\def\be{\begin{equation}}
\def\te{\end{equation}}
\def\bea{\begin{eqnarray}}
\def\nn{\nonumber}
\def\tea{\end{eqnarray}}

\def\ha{{1\over 2}}
\def\tt{\tilde t}
\def\tr{\tilde r}
\def\tS{\tilde S}

\def\t{\tau}

\newskip\humongous \humongous=0pt plus 1000pt minus 1000pt

\newif\ifdtup

\def\ha{{1\over 2}}

\documentstyle[aps,eqsecnum]{revtex} 

\begin{document}

\title{Thermal Particle Creation in Cosmological Spacetimes:
A Stochastic Approach}
\author{
Don Koks \thanks{e-mail address: dkoks@physics.adelaide.edu.au}\\
{\small Department of Physics, University of Adelaide, Adelaide SA 5005,
Australia}\\
B. L. Hu \thanks{e-mail address: hu@umdhep.umd.edu}\\
{\small Department of Physics, University of Maryland,
College Park, MD 20742, USA}\\
Andrew Matacz \thanks{e-mail address: andrewm@maths.su.oz.au}\\
{\small School of Mathematics and Statistics, University of Sydney,
NSW 2006, Australia}\\
Alpan Raval \thanks{e-mail address: raval@csd.uwm.edu}\\
{\small Department of Physics, University of Wisconsin-Milwaukee,
Milwaukee, WI 53201, USA}}
\date{\small {\it (UMDPP 96-116, April 18, 1997)}}
\maketitle

\begin{abstract}
The stochastic method based on the influence functional formalism
introduced in an earlier paper to treat particle creation in 
near-uniformly accelerated detectors and collapsing masses is applied 
here to treat thermal and near-thermal radiance in certain types of 
cosmological expansions. It is indicated how the appearance of thermal 
radiance in different cosmological
spacetimes and in the two apparently distinct classes of black hole and
cosmological spacetimes can be understood under
a unifying conceptual and methodological framework.
\end{abstract}

\newpage
\section{Introduction}

Particle creation in cosmological spacetimes was first discussed by Parker
\cite{Par69}, Sexl and Urbantke \cite{SexUrb69}, Zel'dovich and Starobinsky
\cite{ZelSta71} in the late sixties. The basic mechanism
can be understood as parametric amplification of vacuum fluctuations
by an expanding universe \cite{Zel70,Hu74}.
Particle creation in black hole spacetimes was first discovered by
Hawking \cite{Haw75} (see also \cite{HawRad75}).
A similar effect in a uniformly accelerated
detector was discovered by Unruh \cite{Unr76} and in a moving mirror
by Davies and Fulling \cite{FulDav}. One special class of
cosmological spacetimes which shows this characteristic thermal radiance is
the de Sitter Universe, as shown by Gibbons and Hawking \cite{GibHaw}.
One feature common to all these systems is that they all possess event
horizons,
and the conventional way to understand the thermal character of
particle creation is by way of the periodicity on the propagator of
quantum fields defined on the Euclidean section of the spacetime 
\cite{HarHaw,GibPer}.

One would not usually think of cosmological particle creation as
thermal because in general such conditions (event horizon and periodicity)
do not exist. However, several authors have shown that
thermal radiance  can arise from cosmological particle creation in
spacetimes without an event horizon 
\cite{Ber75,Par76,ChiHar,BerDun,AudSch,BraKha}.
Each case has its particular reason for generating a thermal radiance,
but there is not much discussion of the common ground for these cases
of cosmological spacetimes. There also seems to be a gulf between our
understanding of the mechanisms giving rise to thermal radiance in
these two classes of spacetimes, i.e.,
spacetimes with and without event horizons\footnote{Note that
 both of these  effects are present in particle creation in a non-eternal
 black hole spacetime -- over and above the thermal Hawking radiation
 for an  eternal blackhole, there is also the contribution from backscattering
 of waves over a time-dependent classical effective potential,
 see, e.g. \cite{Page76}.}.

In some of our earlier papers we have discussed thermal radiance
in the class of spacetimes which possess event horizons (uniformly accelerated
detectors \cite{Ang,HM2,RHA}, moving mirrors and black holes \cite{HM2,RHK})
using the viewpoint of exponential scaling \cite{HuEdmonton,cgea} and
the method of statistical field theory \cite{Banff,if}.
In a recent paper \cite{RHK} we show how this method can be applied to
spacetimes which possess event horizons only in some asymptotic
limit, such as near-uniformly or finite-time accelerated detectors,
and collapsing masses. 
In this paper, we study thermal  particle
creation from cosmological spacetimes with the aim of providing a common
ground for cases where thermal radiance was reported before.
Using the stochastic method, we show how to derive near-thermal
radiance in spacetimes without an event horizon.

The two primary examples we picked here for analyzing this issue
are that of Parker and Berger \cite{Ber75,Par76} for an exponentially
expanding universe, and that of slow-roll inflationary universes,
in particular, the Brandenberger-Kahn model \cite{BraKha}.
In Sec.~2 we examine several simple cosmological expansions which
admit thermal particle creation
and show their common ground, and their connection
with thermal radiation in the (static coordinatized) de Sitter spacetime.
We point out that all cases which report thermal radiation involve
an exponential scale transformation \cite{HuEdmonton,cgea}.
Thermal radiance observed in one vacuum
can be understood as arising from the exponential scaling
of vacuum fluctuations of the other vacuum \cite{Dalian}. Sec.~3
applies this method to the Parker-Berger model to show how thermal radiance
can be derived from exponentially-scaled vacuum fluctuations. Sec.~4
discusses particle creation from slow-roll inflationary universe and shows
how near-thermal radiance can be derived in cases which depart from strictly
exponential expansion. We end with a brief discussion in Sec.~5.

\section{Thermal Particle Creation in Cosmological Spacetimes: Exponential
Scaling}

Consider a spatially-flat ($k$=0) Robertson-Walker (RW) universe with metric
\be
ds^2 = dt^2-  a^2 \sum_i ({dx^i})^2,
\te
where $t$ is cosmic time. A conformally-coupled massive ($m$) scalar field
$\Phi$ obeys the wave equation (e.g., \cite{BirDav})
\be
[\Box +m^2 + R/6] \Phi (t, x) = 0,
\te
where $\Box$ is the Laplace-Beltrami operator, and $R= 6 [\ddot a/a + 
(\dot a /a)^2]$ is the curvature scalar. In a spatially-homogeneous space,  
the space and time parts of the wave function separate, with mode decomposition
$ \Phi (t, x) =  \sum_k \phi_k (t) w_k (x) $.  
For a spatially-flat RW universe $w_k (x) = e^{ikx}$, and 
the conformally-related amplitude function $\chi_k (\eta) = a \phi_k (t)$
of the $k^{th}$ mode obeys the wave equation in conformal time 
$\eta = \int dt /a$:
\be
\chi_k (\eta)'' + [k^2 + m^2 a^2(\eta)] \chi_k (\eta) = 0.
\te
Call $\Phi_k^{in,out}(t, x)$ the modes with only positive frequency 
components at $t_-= - \infty$ and $t_+ = + \infty$, respectively.
They are related by the Bogolubov coefficients $\alpha_k, \beta_k$ as 
follows:
\be
\Phi_k^{in} (t, x) = \alpha_k \Phi_k^{out} (t,x) + \beta_k \Phi_{-k}^{out *}
(t,x).
\te
(For conformal fields it is convenient to use the conformally-
related wave function $ X(\eta, x) = a \Phi (t, x)$.
One can define the conformal vacua at $\eta_\pm$ with ${\bf \chi}^{in, out}$
in terms of the positive frequency components.)
The modulus of their ratio is useful for
calculating the probability $P_n(\vec k)$ of observing $n$ particles in mode
$\vec k$ at late times \cite{Par76}:
\be
P_n(\vec k) = |\beta_k /\alpha_k|^{2n} |\alpha_k|^{-2}.
\te
{}From this one can find the average number of particles $\langle N_{\vec k}
\rangle$ created in mode $\vec k$ (in a comoving volume) at late times to be
\be
n_k \equiv \langle N_k\rangle = \sum_{n=0}^\infty n P_n (\vec k) = 
|\beta_k|^2 .
\te

\subsection{Bernard-Duncan Model}

In a model studied by Bernard and Duncan \cite{BerDun} 
the scale factor $a(\eta)$ evolves like
\be\label{case-1}
**~~ {\it Case~1} **  ~~~~~ a^2(\eta) = A + B \tanh \rho \eta  
\te
which tends to  constant values $a^2_{\pm} \equiv A \pm B$ at asymptotic times
$\eta \rightarrow \pm \infty$. Here $\rho$ measures how fast the scale factor
rises, and is the relevant parameter which enters in the temperature of
thermal radiance. With this form for the scale function,
$\alpha_k, \beta_k$ have analytic forms in terms of products of gamma 
functions.  One obtains
\be
|\beta_k/\alpha_k|^2 = \sinh^2 (\pi \omega_-/\rho)/\sinh^2 (\pi 
\omega_+/\rho)
\te
where
\be
\omega_\pm =  \ha (\omega^{out} \pm \omega_{in}), ~~~
\omega^{out}_{in} = \sqrt { k^2 + m^2 a_\pm^2}.
\te
For cosmological models in which $a (+\infty) \gg a (-\infty) $, the argument
of $\sinh$ is very large (i.e. $(\pi/\rho)\omega_{\pm} \gg 1$). To a good 
approximation this has the form
\be\label{about-line-326}
|\beta_k /\alpha_k|^2  = \exp (-2 \pi \omega_{in} /\rho). ~~ 
\te
For high momentum modes, one can recognize the Planckian distribution with 
temperature given by
\be
k_B T_{\eta} = \rho / ( 2 \pi a_+)       
\te
as detected by an observer (here in the conformal vacuum) at late times.

\subsection {Parker-Berger Model}

This model is similar in spirit to the one proposed by Parker earlier
\cite{Par76}, who  considered a massless, minimally coupled scalar field
in a Robertson-Walker universe with metric

\be
ds^2 = a^6 d\tau^2- a^2\sum_i ({dx^i})^2
\te
where $\tau$ is a time defined by $ dt= a^3 d\tau$. The scale factor $a$ is
assumed to approach a constant at $\tau \rightarrow \pm \infty$, where in
these asymptotic regions one can define a vacuum with respect to $\tau$ time
and construct the corresponding field theory. He considered the general 
class of
functions for the scale factor (from Epstein and Eckart \cite{EckEps})
\be\label{case-2}
**~~ {\it Case~2} **  ~~~~a^4 (\tau) = a_1^4 + e^{\sigma \tau} [(a_2^4 - a_1^4)
(e^{\sigma \tau} +1) +b ](e^{\sigma \tau} +1)^{-2}
\te
where $a_1, a_2, b$ are adjustable parameters with $a_2 > a_1$, and $\sigma$
is the rise function (similar to the $\rho$ in the earlier case in terms of the
conformal vacuum). The modulus of the ratio of the  Bogolubov coefficients
is given by \cite{Par76}
\be\label{about-line-359}
\left|\frac{\beta_k}{\alpha_k}\right|^2 = \frac {\sin^2 \pi d + 
\sinh^2 (\pi \omega_-/\sigma)} {\sin^2 \pi d + \sinh^2 (\pi \omega_+/\sigma)}
\te
where $d$ is a real number involving the constant $b$ and
\be
\omega_\pm \equiv k (a_1^2 \pm a_2^2).
\te
For cosmological models in which $a_2 \gg a_1$, the argument of the $\sinh$ 
is very large, as in Case 1. Then, to a good 
approximation~(\ref{about-line-359}) has the form 
\be
|\beta_k /\alpha_k|^2  = \exp (-4 \pi ka_1^2 /\sigma).
\te
This form is independent of the adjustable parameters $b, a_2$.  From this one 
can show that the amount of particle creation is given by
\be
n_k = [\exp (\mu k) - 1]^{-1}
\te
where $\mu = 4 \pi a_1^2 /\sigma$.
Converting to the physical momentum at late times $p= k/a_2$,
one sees that this is a Planckian spectrum characteristic of thermal radiance
with temperature
\be
k_B T_{\tau} = \sigma / (4 \pi a_1^2 a_2).
\te

How sensitively does the thermal character of particle creation  depend
on the scale factor? From physical considerations, the period when particle
creation is significant is when the nonadiabaticity parameter 
$\bar \Omega$ satisfies \cite{Hu74} 
\be
\bar \Omega \equiv  \Omega ' /\Omega^2 \ge 1.
\te
Here $\Omega$ is the natural frequency (given by $ka^2$ in this case)
and $\Omega ' = d \Omega / d\tau$. Using this criterion, Parker argued
that  significant particle creation occurs during an early
period when $e^{\tau/\sigma}$ (or in the first case, $e^{\eta /\rho}$)
is small, whence $a^4$ has effectively the form
\be\label{case-3}
** {\it Case~3} ** ~~~~~~~ a^4(\tau) = a_1^4 + a_0^4 e^{\tau/ \sigma}
\te
where $a_0$ is an adjustable parameter.
This form of the scale factor was used by Berger \cite{Ber75} for the
calculation of particle creation in a Kasner universe, who
also reported thermal radiation. Since particle creation vanishes
at early and late times (as measured by the nonadiabaticity parameter),
adiabatic vacua can be defined then and one can construct WKB positive
and negative frequency solutions for the calculation of the
Bogolubov coefficients. Parker \cite{Par76} showed explicitly that
the modulus of their ratio has the same exponential form as that of the more
complicated scale function~(\ref{case-2}), which, as we have seen, is what
gives rise to the thermal character of the spectrum.
Indeed he speculated that the exponential form in $|\beta /\alpha|$
should hold for a general
class of scale functions which possess the properties that: 1) they
smoothly approach a constant at early time, 2) their values at late times
are much larger than at initial times, and most importantly, that
3) they and their derivatives are continuous functions.
The exponential factor contained in the
scale functions in all three cases above at early times is thus responsible
for the thermal property of particle creation, with the temperature
proportional to the rise factor in the exponential function ($\sigma$, or
$\rho$ in the first case). He also noted that this property
is quite insensitive to the late time asymptotically
static behavior of $a (\t)$  (could be the flattening behavior of a $\tanh$
function, or the rising behavior of an $\exp$ function).

\subsection{Common Features}

The three examples related above highlight an important common feature of
thermal particle creation. That is, that a period of exponential expansion in
the scale factor would give rise to thermal particle creation, i.e.,
an observer in the in vacuum before the expansion reports zero particles,
while an observer in the out vacuum after the exponential expansion will
report a thermal particle spectrum with temperature proportional to the rise
factor in the exponential function in that particular time (e.g.,
conformal time $\eta$ in Bernard and Duncan's model, $\tau$ in Parker's model,
and cosmic time $t$ in the de Sitter
universe example below). The exponential scale factor and the relation between
these two vacua are important to understanding particle
creation in cosmological spacetimes on the same footing as that in the class
of spacetimes with event horizons, including that of a uniform accelerated
observer, a moving mirror, black holes and the de Sitter universe.

To put the physics in a more general context, consider two vacua related
by some transformation. Let us define the asymptotic
in-vacuum as $|0\rangle_t$, the asymptotic out-vacuum as $|0\rangle_{out}$
and the vacuum of an observer undergoing exponential expansion
as $|0\rangle_s$. The in and out vacua are well defined because the scale 
factor 
approaches a constant at asymptotic past and future times, thus imparting the
space with a Killing vector $\partial_t$ with respect to which one can define
particle states in terms  positive frequency modes. The $s$-vacuum is
defined with respect to a different set of mode functions (like the Fulling-
Rindler vacuum vis-a-vis the Minkowski vacuum for a uniformly- accelerated
observer). The above examples calculate the particle creation between an in
and out vacuum, but they also illustrate the important fact that the number of
particles created is insensitive to the behavior of the scale function at
late times -- e.g., the result for the flattening $\tanh$ function which gives
an asymptotically static universe is the same as that of the exponential
function.
Furthermore, the thermal nature of particle creation depends only on the
initial stage of exponential expansion. In \cite{Dalian,HuRav96}
these findings were used to connect the result of thermal radiance in these
two classes of spacetimes. The assertion is that the more basic cause of
thermality lies in the exponential scaling behavior rather than the existence
of an event horizon \cite{HuEdmonton,cgea}. (The latter necessarily implies the
former, but the converse is not always true). The fine distinction
between these two ways of understanding (kinematic versus geometric)
thermal radiance will enable us to treat non-thermal cases
in spacetimes which do not possess event horizons \cite{RHK},
and to explore the stochastic nature of the Hawking-Unruh effect.

%

At this point it is perhaps also useful for us to adopt this kinematic 
viewpoint to reexamine the cause of thermal radiance in a de Sitter universe.

\subsection{de Sitter Universe}

The de Sitter universe metric can be expressed in many different coordinates
(see, e.g. \cite{BirDav}). There is the so-called closed ($k$=1) 
Robertson-Walker 
(RW) coordinatization which covers the whole space, $a(t) = \cosh (Ht) $, 
the flat ($k$=0) RW coordinatization which covers only half of de Sitter
space with scale factor $a (t) = e^{Ht}$, and the static coordinate shown 
below, to name just the commonly encountered ones.
The vacuum states defined with respect to different coordinatization
and normal mode decomposition have been studied by many authors
\cite{vacds}. In the static coordinate the metric is given by
\be
ds^2 = [1- (H\tr)^2] d \tt^2 - {{d \tr^2 }\over {[1- (H\tr)^2 ]}}.
\te
Note that an event horizon exists at $\tr = H^{-1}$ for observers at $\tr = 0$,
following the trajectory generated by the Killing vector $\partial_{\tt}$.
This is similar in form to a Schwarzschild metric for a massive ($M$) 
object, restricted to 2 dimensions:
\be
ds^2 = (1- {{2M}\over{r}}) dt^2 - {{dr^2}\over{(1- {{2M}\over{r}})}},
\te
for which Hawking \cite{Haw75} first reported the famous black hole thermal
radiance effect.

Calculation of particle creation in the static de Sitter universe was carried
out by Gibbons and Hawking \cite{GibHaw} (GH) using the periodicity 
condition in the field propagator.  Lapedes \cite{Lap} gave derivations 
based on the use of Bogolubov transformations, and Brandenberger and Kahn 
\cite{BraKha} treated the asymptotically de Sitter case. 
Let us analyze the relation
between  particle creation calculated in the Gibbons-Hawking vacuum
(defined with respect to the `static' de Sitter time $\tt$)  
$|0\rangle_{\tt}$ and that in the Robertson-Walker ($k$=0) vacuum
(defined with respect to cosmic time $t$) $|0\rangle_t$.
We will see that the $|0\rangle_{\tt}$ vacuum bears
the same relation to $|0\rangle_t$ vacuum as that between the exponential 
vacuum $|0\rangle_s$ defined earlier and the asymptotic in-vacuum 
$|0\rangle_t$ in the cosmological cases above.


\subsection{Exponential scaling:  a kinematic viewpoint}

Starting from special relativity, assuming that two coordinate systems 
$ S = (t, r)$ and  $ \tS = (\tt, \tr)$ (these are not the black 
hole coordinates) coincide at the origin, so
the initial vacuum of $\tS$ is the same as the RW vacuum, but the final vacuum
in $\tS$ is the GH vacuum. The two systems are related by the following 
conditions:
\be
(i) ~~~ \tr = a(t) r
\te
\be
(ii)~~~ a(t) = e^{Ht}
\te
\be
(iii)~~~ H\tr = Har = \dot a r = \beta,
\te
\be
(iv)~~~ {{a(\tt)} \over {a(t)}} = \gamma = {1 \over {\sqrt{1 - \beta^2}}}
\te
The meaning of these conditions is explained in \cite{Dalian}, which uses
this example to illustrate the existence of exponential scaling in all
cases which report thermal radiance. The two systems are related by a scale
 transformation {\it (i)}, in this case, an
exponential scaling {\it (ii)} such that an observer in $\tS$ is seen as 
receding
from $S$ with a velocity of $\beta$, with $H$ the red-shift or Hubble parameter
{\it (iii)}, and a Lorentz factor $\gamma$ {\it (iv)}. Condition {\it 
(iv)} is called relativistic exponential  transformation \cite{Dalian},
which plays a central role in the understanding of the Hawking effect in terms 
of scaling concepts \cite{HuEdmonton,cgea,KM}.
With these correspondences, it is not difficult to see the analogy with the
cosmological particle creation cases studied before. The initial (asymptotic 
in-) vacuum $|0\rangle_t$ defined with respect to  $t$ time here (or $\eta, 
\tau$ time in the
earlier examples) and the vacuum  $|0\rangle_{\tt}$ defined in the 
exponentially
receding system $\tS$ bear the same relation. It is no surprise that the
modulus of the ratio between the Bogolubov coefficients have the same
form characteristic of a thermal spectrum~(\ref{about-line-326}), but with
$\rho, \sigma$ replaced by $H$. [One can find explicit
calculation of thermal particle creation in Eq. (33) of \cite{Lap},
using Bogolubov transformations replacing $R$ by our $H^{-1}$.] 
The Hawking temperature for the de Sitter universe is given by
\be
k_B T_{dS} = \frac{H}{2 \pi}.
\te

Once the relation between
the de Sitter universe (which belongs to the class of spacetimes which
show the Hawking-Unruh effect), and that of some general cosmological 
spacetimes (with specific scale functions such as in~(\ref{case-1},
\ref{case-2}, \ref{case-3}) is established, it is easy to
generalize to the black hole and accelerated detector cases.

\section{Thermal Radiance in the Parker-Berger Model}

We now use the influence functional formalism (see e.g., \cite{HM2})
to investigate particle creation in the Parker-Berger 
model~\cite{Ber75,Par76}. 
The line element 
is given by
\be
ds^2 = a^6 d\tau^2- a^2\sum_i ({dx^i})^2
\te
where 
\be
a^4(\tau) = 1+e^{\rho\tau}.
\te
We consider the action of a massless, minimally coupled real scalar field 
$\phi$, which forms an environment acting upon a detector coupled to this 
field at some point in space.  The field can be decomposed into a collection of
oscillators of time-dependent frequency.  Using the influence functional 
formalism, we can determine the
effect of such an environment on the detector, which is also modelled by an
oscillator.

To do this we calculate the noise $\nu$ and dissipation $\mu$ produced by the
field.  These are given by:
\be\label{blue32.4-1}
\zeta \equiv \nu + i\mu = \int_0^\infty dk\;I(k,s,s')X(s)X^\ast(s')
\te
where $I$ is the spectral density describing the system/environment 
interaction, and $X$ is a sum of Bogolubov coefficients satisfying the 
classical equation of motion for the field oscillators.
First we decompose the field into its modes; the Lagrangian density is
\bea
{\cal L}(x) &=& {\sqrt{-g}\over 2} \phi^{,\mu}\phi_{,\mu} \nn \\
&=& {1\over 2}\left[\phi_{,\tau}^2 - a^4 \sum_i 
\left(\phi_{,i}\right)^2\right]. 
\tea
In terms of normal modes the Lagrangian becomes
\be
L(\tau) = \sum_{k,\sigma=\pm} {1\over 2} \left[({q_{\bf k}^\sigma}_{,\tau})^2-
a^4k^2({q_{\bf k}^\sigma})^2\right].
\te
We see then that the bath can be described as a set of oscillators with mass 
and frequency
\be
m = 1 \;\;\;,\;\;\; \omega^2 = a^4k^2.
\te
Now as was already mentioned, $X$ satisfies the classical equation of motion
for an oscillator with the given parameters, and being a sum of Bogolubov
coefficients, its initial values are predetermined.  So we need to solve
\be\label{1-1}
X''(\tau) + k^2(1+e^{\rho\tau}) X = 0,
\te
with initial conditions:
\be
X(\tau_0) = 1 \;\;\;,\;\;\; X'(\tau_0) = -ik.
\te
With a change of variables $z = \ln (2k/\rho) + \rho\tau/2$ we find
the solutions in terms of the Bessel functions:
\be
X(\tau) = c_1 J_{2ik\over\rho} \left({2k\over\rho}e^{\rho\tau/2}\right)+
          c_2 J_{-{2ik\over\rho}}\left({2k\over\rho}e^{\rho\tau/2}\right)
\te
To fix the constants $c_1,c_2$ consider that the initial time is 
$\tau_0\rightarrow -\infty$; however the complex index Bessel functions
oscillate infinitely often as their arguments approach zero, and so for now we
leave~$\tau_0$ unspecified.  In that case we can calculate%
\footnote{To calculate
these coefficients the wronskian of $J_{2ik/\rho}$ and $J_{-2ik/\rho}$ is
needed; note that there is a misprint in Gradshteyn and Ryzhik \S 8.474: 
the relevant quantity should be $-{2\over\pi z}\sin\nu\pi$.}
$c_1$ and $c_2$; the final expression for $X$ becomes, with
$$
f(\tau) \equiv {2k\over\rho}e^{\rho\tau/2},
$$
and Bessel indices labelled by $\nu \equiv 2ik/\rho$:
\be
X(\tau) = {\pi k\over\rho}{\rm csch}{2\pi k\over\rho}\left\{
ie^{\rho\tau_0/2}\left|
\begin{array}{cc}
   J_\nu(f(\tau)) & J_{-\nu}(f(\tau)) \\
   J_\nu'(f(\tau_0)) & J_{-\nu}'(f(\tau_0))
\end{array}\right|
-\left|
\begin{array}{cc}
   J_\nu(f(\tau)) & J_{-\nu}(f(\tau)) \\
   J_\nu(f(\tau_0)) & J_{-\nu}(f(\tau_0))
\end{array}\right|\right\}.
\te
In the limiting case of $\tau\rightarrow\infty$ ($\tau_0\rightarrow -\infty$)
we can use first order and asymptotic
expressions for $J$ and $J'$ to write (which defines the phases $\alpha$ and 
$\beta$)
\bea
J_\nu(f(\tau_0)) &\simeq& \sqrt{\sinh 2\pi k/\rho \over 2\pi k/\rho} 
e^{i\alpha(k)}\nn \\
J_\nu'(f(\tau_0)) &\simeq& e^{-\rho\tau_0/2} \sqrt{\sinh 2\pi k/\rho \over 
2\pi k/\rho} e^{i\beta(k)} \nn \\
J_\nu(f(\tau)) &\simeq& \sqrt{2\over \pi z} \cos \left(f(\tau)-{\nu\pi\over 2}-
{\pi\over 4}\right).
\tea
In evaluating $X(\tau)X^\ast(\tau')$ we obtain various products of the Bessel
functions with
their derivatives (note: $J_\nu^\ast = J_{-\nu}$); in particular we need
\bea\label{12-1}
\beta - \alpha &=& {\rm arg}\;\Gamma\left(1+{2ik\over\rho}\right) -
                   {\rm arg}\;\Gamma\left(2ik\over\rho\right) \nn \\
&=& {\rm arg}\;{2ik\over\rho} = {\pi\over 2} \;\;{\rm provided}\;k\ne 0.
\tea
Also, when calculating the Bessel products, there arise sines and cosines
with argument $f(\tau)+f(\tau') \equiv 2k/\rho \left(e^{\rho \tau/2}+
e^{\rho \tau'/2} \right)$; when $\tau\rightarrow\infty$ and we ultimately 
integrate over $k$,
these terms won't contribute to the integral and so can be discarded.  
Changing to sum and difference variables defined by
\be
\Sigma \equiv (\tau+\tau')/2 \;\;\;,\;\;\;
\Delta \equiv \tau-\tau'
\te
we finally obtain
\be\label{parker15-1}
\zeta = e^{-\rho\Sigma/2}\int_0^\infty dk\;I(k,\tau,\tau')
\left[\cos{2k\over\rho}
\left(e^{\rho \tau/2}-e^{\rho \tau'/2}\right)\coth{2\pi k\over\rho}-i\sin
{2k\over\rho}\left(e^{\rho \tau/2}-e^{\rho \tau'/2}\right)\right].
\te
We can now equate $\zeta$ with the standard form:
\be\label{16-1}
\zeta = \int_0^\infty dk\;I_{\it eff}(k,\Sigma)\;\left[C(k,\Sigma)
\cos k\Delta-i\sin k\Delta\right].
\te
This is the form of $\zeta$ which, with the function $C$ replaced by 
$\coth {k\over 2T}$, would describe a thermal bath of static 
oscillators each in a coherent state.  We will show that the
unknown function $C$ does indeed have the form of a coth, and can then
deduce the temperature of the radiation seen by the detector. Here
$I_{\it eff}(k,\Sigma)$ is the effective spectral density, also to  
be determined.  We can always write $\zeta$ in this way since $\nu$ is even in 
$\Delta$ while $\mu$ is odd.  By equating the real and imaginary parts of the 
two forms of $\zeta$ and Fourier inverting, we obtain
\bea\label{bk-10.05-1-and-dec-5-7-and-bk-10.1-1}
I_{\it eff}C &=& {1\over\pi}
\int_{-\infty}^\infty d\Delta\;\cos k\Delta\;\nu(\Sigma,\Delta),\nn \\
&&\nn \\
I_{\it eff} &=& -{1\over \pi}
\int_{-\infty}^\infty d\Delta\;\sin k\Delta\;\mu(\Sigma,\Delta).
\tea
These expressions will be used throughout this paper to calculate $C$ for the
various cases of induced radiance that we consider.  In order to use
these we need to calculate the dissipation and noise, $\mu$ and $\nu$.

We first evaluate $\mu$ as given by (\ref{parker15-1}); substituting it 
into~(\ref{bk-10.05-1-and-dec-5-7-and-bk-10.1-1}) will then give us the 
effective spectral density $I_{\it eff}(k,\Sigma)$. Define
\be
\sigma \;\equiv \;{2\over\rho}\left(e^{\rho s/2}-e^{\rho s'/2}\right)
\;= \;{4\over\rho} e^{\rho\Sigma/2}\sinh{\rho\Delta\over 4}.
\te
To proceed, we need to specify a form for the spectral density.  This has been
calculated in~\cite{HM2}, and in 3+1 dimensions it is
\be
I(k,s,s') = {c^2 k\over 4\pi^2},
\te
where $c$ is the coupling strength of the detector to the field.  Then 
from~(\ref{parker15-1}) we have
\bea
\mu &=& -{c^2\over 4\pi^2}e^{-\rho\Sigma/2}\int_0^\infty dk\; k\sin\sigma k 
\nn\\
&=& {c^2\over 4\pi^2}e^{-\rho\Sigma/2}\;\pi\delta'(\sigma)
\tea
where the last result follows from the discussion in \cite{KMH}.
Substituting this form for $\mu$ 
into~(\ref{bk-10.05-1-and-dec-5-7-and-bk-10.1-1}) gives the following result:
\be
I_{\it eff}(k,\Sigma) = {c^2k\over 4\pi^2} e^{-3\rho\Sigma/2}.
\te

Evaluating the noise kernel $\nu$ is a more complicated affair.
From~(\ref{parker15-1}) we write
\bea
\nu &=& {c^2\over 4\pi^2}e^{-\rho\Sigma/2}\int_0^\infty dk\; k\cos\sigma k 
\coth{2\pi k\over\rho} \nn\\
&& \nn\\
&=& {c^2\over 4\pi^2}e^{-\rho\Sigma/2}\left[{d\over d\sigma}P(1/\sigma) +
{1\over\sigma^2}-{\rho^2\over 16}{\rm csch}^2{\rho\sigma\over 4}\right]
\tea
where again the last integral has been calculated in \cite{KMH}.
Upon substituting this into~(\ref{bk-10.05-1-and-dec-5-7-and-bk-10.1-1})
we obtain
\bea\label{18-2}
C(k,\Sigma) &=& {e^{\rho\Sigma}\over\pi k}\int_{-\infty}^\infty d\Delta\;
\cos k\Delta {d\over d\sigma}P(1/\sigma) \nn \\
&&{} + {\rho^2\over 16\pi k}\int_{-\infty}^\infty
d\Delta\;\cos k\Delta \left[{\rm csch}^2\;{\rho\Delta\over 4}-e^{\rho\Sigma}
{\rm csch}^2\left(e^{\rho\Sigma/2}\sinh{\rho\Delta\over 4}\right)\right].
\tea
The first integral can be done by parts to get
\bea
\int_{-\infty}^\infty d\Delta\;\cos k\Delta {d\over d\sigma}P(1/\sigma) &=&
\int_{-\infty}^\infty d\sigma\;{d\Delta\over d\sigma}\cos k\Delta {d\over 
d\sigma}P(1/\sigma) \nn \\
&=& -{\rm PV}\int_{-\infty}^\infty {d\Delta\over \sigma}{d\over d\Delta}
\left[{d\Delta\over d\sigma}\cos k\Delta\right] \nn \\
&=& {4\pi k\over\rho}\coth{2\pi k\over\rho}.
\tea
The second integral in (\ref{18-2}) does not appear to be expressible in terms 
of known functions.  Suppose we call it $B(k,\rho,\Sigma)$, and write
\be
B(k,\rho,\Sigma) = 2\int_0^\infty d\Delta\;\cos k\Delta \left[{\rm csch}^2\;
{\rho\Delta\over 4}-e^{\rho\Sigma}{\rm csch}^2\left(e^{\rho\Sigma/2}
\sinh{\rho\Delta\over 4}\right)\right].
\te
Then we have
\be
C(k,\rho,\Sigma) = e^{\rho\Sigma}\left[{4\over\rho}\coth{2\pi k\over\rho}
+ {\rho^2\over 16\pi k}e^{-\rho\Sigma}B(k,\rho,\Sigma)\right].
\te
We need to examine the second term in the last brackets.
The function $B$ tends to zero for large $k$ (by the Riemann-Lebesgue lemma),
and attains
a maximum at $k=0$ (since the cos term stops oscillating there).  However,
numerical work shows that this maximum value increases roughly with 
$e^{\rho\Sigma}$ which means that on first glance the second term in the
brackets does not necessarily vanish at late times 
($\Sigma\rightarrow\infty$).
So we need to examine the value of $B$ at $k=0$ more closely, to see
precisely how it changes with $\Sigma$.  To this 
end we can consider $B(0,\rho,\Sigma)$ as a function of $e^{\rho\Sigma}$ 
and analyze its concavity; i.e.\ with $x\equiv e^{\rho\Sigma}$ we need 
$\partial^2B/\partial x^2$.  Differentiating twice under the integral sign 
gives an integrand which is everywhere negative, and so we conclude that 
$\partial^2B/\partial x^2 < 0$, which means that $B$ as
a function of $x$ is everywhere concave down.  But $B$ increases with $x$, and
thus $B/x\equiv e^{-\rho\Sigma}B\rightarrow 0$ as $\Sigma\rightarrow\infty$.
In that case the second term in the brackets gives no contribution in the large
time limit.

Finally, from~(\ref{16-1}) we can write $\zeta$ in a form
which reveals the thermal nature of the detected radiation:
\be
\zeta =\int_0^\infty dk\;I_{\it eff}(k,\Sigma)\;\left[
{4e^{\rho\Sigma}\over\rho}\coth{2\pi k\over\rho}
\cos k\Delta-i\sin k\Delta\right].
\te
The temperature of the radiation is then
\be\label{20.6-1}
k_BT = {\rho \over 4\pi}.
\te
%
%
Inspection of (\ref{parker15-1}) suggests that an alternative time variable can
be chosen:
\be
t = {2\over\rho} e^{\rho\tau/2}.
\te
The metric becomes
\be
ds^2 = {4a^6\over \rho^2 t^2}dt^2- a^2\sum_i ({dx^i})^2
\te
with
\be
a^4 = 1+{\rho^2 t^2\over 4}.
\te
Again following the previous formalism, we arrive at a description of
the environment field in terms of oscillators, now with time dependent
mass and frequency:
\be
m = {\rho t\over 2} \;\;\;,\;\;\; \omega^2 = {4a^4k^2\over \rho^2 t^2}.
\te
Solutions for $X$ in this case are the same as before, and the calculations 
carry through in much the same way.  With now $\Sigma$ and $\Delta$ defined as
mean and differences of $t$ and $t'$ we again arrive at thermal forms for
the noise and dissipation:
\be
\zeta = e^{-\rho\Sigma/2}\int_0^\infty dk\;I(k,s,s')
\left[\cos k\Delta\coth{2\pi k\over\rho}-
i\sin k\Delta\right]
\te
and the detected temperature is the same as (\ref{20.6-1}).

\section{Inflationary Universe}

\subsection{Eternal versus Slow-roll Inflation}


In this section we consider particle creation of a massless conformally coupled
quantum scalar field at zero temperature in a spatially flat FRW universe 
undergoing a near-exponential (inflationary) expansion.
The example of de~Sitter space which corresponds to the exact exponential case
has been treated in~\cite{HM2}.
Here we first
solve for a general scale factor $a(t)$ using a slightly different language
from~\cite{HM2}.  We then specialise to a spacetime (the Brandenberger-Kahn
metric \cite{BraKha}) which has initial de Sitter behavior but with scale
factor tending toward a constant at late (cosmic) times.  We can also define 
a parameter $h$ which measures the departure from an exact exponential 
expansion.

As before we first derive the noise and dissipation kernels
by calculating $X$, the solution to the equation of motion of the field modes.
The spatially-flat FRW metric is
\be
ds^2 = dt^2-a^2(t)\sum_i {(dx^i)}^2.
\te
The Lagrangian density of the scalar field is
\be\label{3.631.6-1}
{\cal L}(x) = {a^3 \over 2}\left[\dot\Phi^2-{1\over a^2}\sum_i \Phi_{,i}^2
- \left({\dot a^2\over a^2}+{\ddot a\over a}\right)\Phi^2\right]
\te
which leads to a Lagrangian in terms of the normal modes $q_{\bf k}$
\be\label{3.631.6-6}
L(t) = \sum_{k,\sigma=\pm} {a^3\over 2} \left[{(\dot q_{\bf k}^\sigma)}^2
+2{\dot a\over a}\dot q_{\bf k}^\sigma q_{\bf k}^\sigma
-\left({k^2\over a^2}-{\dot a^2\over a^2}\right){(q_{\bf 
k}^\sigma)}^2\right]. \te
(We have added a surface term $1/a^3 \;d/dt\; (\dot a a^2q^2)$ to the 
Lagrangian. See \cite{HM2} for the rationale.)
The classical equation of motion, and hence that of $X$, is
\be
\ddot X + 3{\dot a\over a} \dot X + \left({k^2\over a^2}+{\dot a^2\over a^2}
+{\ddot a\over a}\right) X = 0
\te
with initial conditions
\be
X(t_i) = 1 \;\;\; , \;\;\; X'(t_i) = -ik - a'(t_i).
\te
We find that
\be
X(t) = {1\over a}e^{-ik\eta}
\te
where $\eta$ is the usual conformal time.

Now we use (\ref{blue32.4-1}) to construct the influence kernel~$\zeta$, which
contains the noise and dissipation kernels.
From~\cite{HM2} the spectral density for the field is
\be
I(k,t,t') = {\varepsilon^2 k\over 4\pi^2}.
\te
For the rest of this section we introduce for brevity:
\bea\label{15.1-2}
x(\Sigma, \Delta) &\equiv& a(t) + a(t'), \nn\\
y(\Sigma, \Delta) &\equiv& \eta(t) - \eta(t').
\tea
Then we find from (\ref{blue32.4-1})
\bea\label{5.7}
\zeta \equiv \nu + i\mu &=& {1\over a(t) a(t')} 
{\varepsilon^2\over 4 \pi^2} \int_0^\infty  k\;e^{-iky}\;dk \nn\\
&=& {1\over a(t) a(t')}{\varepsilon^2\over 4 \pi^2} \left[{d\over dy}P(1/y)
+ i \pi \delta'(y)\right].
\tea
As before we calculate the spectrum and temperature by
Fourier transforming $\zeta$ 
using~(\ref{bk-10.05-1-and-dec-5-7-and-bk-10.1-1}).
(Note that $\Sigma,\Delta$ are defined in terms of cosmic time $t$).
We change the variable of integration from $\Delta$ to $y$, using
\be\label{10.21-2.5}
{d\Delta\over dy} = {2a(t)a(t')\over x}
\te
to write
\bea\label{10.1-2}
I_{\it eff} &=& -{\varepsilon^2 \over 2\pi^2}\int_{-\infty}^\infty \delta'(y)
{\sin k\Delta\over x}\;dy \nn\\
&=& {\varepsilon^2 k\over 4\pi^2}
\tea
independently of the value of $a(t)$.
The temperature now follows from~(\ref{bk-10.05-1-and-dec-5-7-and-bk-10.1-1}):
again we change to a $y$-integration by parts, remembering that 
$\Delta = 0 \Longleftrightarrow y=0$.  We obtain
\be\label{10.05-0.5}
I_{\it eff}C = {\varepsilon^2 \over 4\pi^3}\int_{-\infty}^\infty{\cos k\Delta
\over a(t)a(t')}{d\Delta\over dy}{d\over dy}P(1/y)\;dy
\te
which leads to
\be\label{10.05-5}
C = -{4\over\pi k}\int_0^\infty\left[{d\over d\Delta}\;{\cos k\Delta\over x}
\right]{d\Delta\over y}. 
\te
This equation is the central result of this section, in that it allows us to
compute the spectrum corresponding to an arbitrary scale factor.
For example, in the de Sitter case with $a = e^{Ht}$ we have
\be\label{15.2-3-and-15.4-4}
x = 2e^{H\Sigma}\cosh {H\Delta\over 2} \;\;\;;\;\;\;
y = {2e^{-H\Sigma}\over H}\sinh {H\Delta\over 2}
\te
which when substituted into (\ref{10.05-5}) gives
\be\label{10.2-5}
C = \coth {\pi k\over H}.
\te
So for this case we can infer the temperature seen to be
\be
k_BT = {H\over 2\pi}
\te
as was calculated with a slightly different approach in \cite{HM2}.

As an aside, we note that from the above analysis for a general scale factor, 
the noise kernel is
\be\label{10.3-1}
\nu = -{\varepsilon^2 \over \pi^3}\int_0^\infty dk\cos k\Delta\int_0^\infty 
\frac{du}{y}{{d\over du}\left[{\cos ku \over a(t)+a(t')}\right]}
\te
with dissipation
\be\label{10.3-2}
\mu = {\varepsilon^2 \delta'(\Delta) \over 4\pi}.
\te
An often-used alternative to our principal part prescription is the 
introduction of a cutoff in the $\int_0^\infty$ expressions; unfortunately 
following this procedure doesn't lead to 
tractable integrals even for the relatively simple de Sitter case.
Note that in equation~(\ref{10.05-5}) for the temperature in the general 
case, we are essentially dealing with products of $a$ and
$\eta$, and it's therefore not surprising that for de Sitter expansion, where
$a \propto 1/\eta$, that (\ref{10.05-5}) can be evaluated analytically.  For 
other forms of $a$, even very simple ones, (\ref{10.05-5}) becomes very 
complicated.

\subsection{Near-exponential expansion}

We now consider the case of a near de Sitter universe with a
scale factor composed of the usual de Sitter one together with a factor
that decays exponentially.  We show that the spectrum seen is near-thermal
tending toward thermal at late times.

We start by considering the Hubble parameter to have a constant value 
(characterising de Sitter space) plus an exponentially decaying term:
\be\label{15-1}
H(t) = H_0\left(1+\alpha e^{-\beta H_0 t}\right),
\te
and from this our aim is to calculate $C$ using (\ref{10.05-5}).  The scale
factor results from integrating $H$, and is
\be\label{15-2}
a(t) = \exp\left(H_0t-{\alpha\over\beta}e^{-\beta H_0 t}\right).
\te
We define the parameter $h$ which measures the departure from exact 
exponential expansion to be
\be
h(t) \equiv \frac{\dot H(t)}{H(t)^2} \to -\alpha\beta e^{-\beta H_0 t}
\te
as $\beta t \to \infty$, and as we might expect it is exponentially decaying
at late times.

To proceed, we indicate the de Sitter quantities by a subscript zero as well
as writing 
\be
\widetilde \Sigma \equiv H_0 \Sigma \;\;\; , \;\;\;
\widetilde \Delta \equiv H_0 \Delta.
\te
Then immediately we have from (\ref{15.2-3-and-15.4-4}):
\be
x_0 = 2e^{\widetilde\Sigma}\cosh {\widetilde\Delta\over 2} \;\;\;;\;\;\;
y_0 = {2e^{-\widetilde\Sigma}\over H}\sinh {\widetilde\Delta\over 2}.
\te
We wish to perturb these by using the new scale factor.  Suppose we define
perturbations $f_1, f_2$ by writing
\be
x = x_0 \Bigl(1+f_1(\widetilde\Sigma, \widetilde\Delta)\Bigr) \;\;\;,\;\;\;
y = y_0 \Bigl(1+f_2(\widetilde\Sigma, \widetilde\Delta)\Bigr).
\te
We first have
\be\label{15.2.5-0}
x = a(t) + a(t') = e^{H_0t-{\alpha\over\beta}e^{-\beta H_0 t}}+
e^{H_0t'-{\alpha\over\beta}e^{-\beta H_0 t'}},
\te
which in the late time limit can be approximated by
\be\label{15.2.5-1}
x \simeq x_0 -{2\alpha\over\beta}\;e^{(1-\beta)\widetilde\Sigma}\;
\cosh{(1-\beta)\widetilde\Delta\over 2},
\te
which yields $f_1$:
\be\label{15.2.5-2}
f_1 = {-\alpha\over\beta}\;e^{-\beta\widetilde\Sigma}\;{\cosh{(1-\beta)
\widetilde\Delta\over 2}\over \cosh{\widetilde\Delta\over 2}}.
\te
Next we write
\bea\label{15.3-6}
y = \eta(t) - \eta(t') &=& \int_{t'}^t {dt\over a(t)} \nn\\
&=& \int_{t'}^t \exp\left(-H_0t+{\alpha\over\beta}e^{-\beta H_0 t}\right),
\tea
and by making the same late time approximation as for $x$ we get
\bea\label{15.4}
y &\simeq& \int_{t'}^t e^{-H_0t}\left(1+{\alpha\over\beta}e^{-\beta H_0 t}
\right) \nn\\
&=& y_0 + {2\alpha \;e^{-(1+\beta)\widetilde\Sigma}\over \beta(1+\beta)H_0}
\sinh{(1+\beta)\widetilde\Delta\over 2}.
\tea
This leads to
\be\label{15.4-5}
f_2 = {\alpha\;e^{-\beta\widetilde\Sigma}\over \beta(1+\beta)}{\sinh{(1+\beta)
\widetilde\Delta\over 2}\over \sinh{\widetilde\Delta\over 2}}.
\te
Note that at late times $f_1, f_2$ tend to zero.  In that case to calculate $C$
we write~(\ref{10.05-5}) in the form
\be\label{15.5-1}
C \simeq -{4\over \pi k}\int_0^\infty{d\over d\Delta}\left[{\cos k\Delta\over
x_0}(1-f_1)\right]{1-f_2\over y_0}\;d\Delta, 
\te
and so write, to first order in $f_1, f_2$:
\be\label{15.5-5}
C \simeq \underbrace{{-4\over \pi k}\int_0^\infty{d\Delta\over y_0}{d\over
d\Delta}{\cos k\Delta\over x_0}}_{=\coth\pi k/H_0} 
\;\;+\;\; \underbrace{{4\over \pi k}\int_0^\infty {d\Delta\over y_0}
\left[f_2 {d\over d\Delta}{\cos k\Delta\over x_0}+{d\over d\Delta}{f_1\cos k
\Delta\over x_0}\right]}_{\equiv \Delta C,\mbox{\rm \ the perturbation}}.
\te
Evaluating $\Delta C$ is lengthy but straightforward so we merely write the 
answer in integral form:
\bea\label{15.7-2}
\Delta C &=& {H_0^2\alpha\;e^{-\beta\widetilde\Sigma}\over 2\pi k\beta}\;
\int_0^\infty d\Delta\left[{-2k\over H_0(1+\beta)}{\sinh{(1+\beta)\widetilde
\Delta\over 2}\; \sin k\Delta\over \sinh^2{\widetilde\Delta\over 2} 
\cosh{\widetilde\Delta\over 2}}
-{1\over 1+\beta}{\sinh{(1+\beta)\widetilde\Delta\over 2}\;\cos k\Delta\over
\cosh^2{\widetilde\Delta\over 2} \sinh{\widetilde\Delta\over 2}} \right.\nn\\
&&\nn\\
&&\left.{}+ {2k\over H_0}{\cosh{(1-\beta)\widetilde\Delta\over 2} \sin k\Delta
\over\cosh^2{\widetilde\Delta\over 2} \sinh{\widetilde\Delta\over 2}}
- {(1-\beta)\sinh{(1-\beta)\widetilde\Delta\over 2} \cos k\Delta\over \cosh^2
{\widetilde\Delta\over 2} \sinh{\widetilde\Delta\over 2}}
+ {2 \cosh{(1-\beta)\widetilde\Delta\over 2} \cos k\Delta\over \cosh^3
{\widetilde\Delta\over 2}}\right]. \nn\\&&
\tea
The important point is that the factor $e^{-\beta\widetilde\Sigma}$ ensures 
that this perturbation to the thermal spectrum dies off exponentially at late
times.

\subsection{Brandenberger-Kahn model}

We are now in a position to derive the function $C(k,\Sigma)$ for the 
Brandenberger-Kahn model.  In this case,
\be
a(t) = e^{ {2H_0\over\alpha}\left(1-e^{-\alpha t/2}\right)}
\te
with $H_0,\alpha$ constants.  As $t$ tends toward zero and infinity, $a(t)$ 
tends toward $e^{Ht}$ and $e^{2H/\alpha}$ respectively.  The Hubble expansion 
function is
\be
H(t) \equiv \frac{\dot a}{a} = H_0 e^{-\alpha t/2}
\te
and the parameter which measures the departure from exact exponential expansion
$h(t)$ is
\be\label{bk-12.13.1-0.5}
h \equiv \frac{\dot H(t)}{H(t)^2} = - \frac{\alpha}{2H_0} e^{\alpha t/2}
= -{\alpha\over 2 H_0} + O(\alpha^2t^2).
\te
We assume that $|\alpha t| \ll 1$.
Eqn~(\ref{10.05-5}) is much too difficult to evaluate analytically here, but we
can get some insight by calculating it as a first order correction in $h$
to the de Sitter case.

At this point, we also mention an alternative perturbation of de Sitter
space, given by the scale factor
\be
a(t) = e^{\int_{0}^{t} H(t)dt}
\te
which describes a solution of the vacuum Einstein equations with a 
time-dependent cosmological constant $\Lambda(t) = 3H^2(t)$. One may expand
$H(t)$ in a power series about $t=0$. Defining $h$ as
in~(\ref{bk-12.13.1-0.5}),
this form of perturbation turns out to be identical to the Brandenberger-Kahn
model to first order in $h$. We have, to first order,
\be
H(t) = H_0 + H_{0}^{2} h t,
\te
the correspondence between $h$ and $\alpha $ being given
by~(\ref{bk-12.13.1-0.5}).
We will therefore calculate the detector response for the Brandenberger-Kahn
model only, keeping in mind its correspondence with the model mentioned above.

Again define $f_1, f_2$ (now with $h$ included) such that
\be
x = x_0 + h f_1(\Delta) \;\;\;,\;\;\;
y = y_0 + h f_2(\Delta).
\te
The corrections are then written as
\bea\label{12.13.1-3}
f_1(\Delta) &=& e^{\widetilde\Sigma}\left[\left(\widetilde\Sigma^2+{\widetilde
\Delta^2\over 4}\right)\cosh{\widetilde\Delta\over 2} + \widetilde\Sigma
\widetilde\Delta\sinh{\widetilde\Delta\over 2}\right],\nn\\
f_2(\Delta) &=& -{e^{-\widetilde\Sigma}\over 
H_0}\left[\left(\widetilde\Sigma^2 +{\widetilde\Delta^2\over 4}+
2\widetilde\Sigma+2\right)\sinh {\widetilde\Delta\over 2} -
\left(\widetilde\Sigma+1\right)\widetilde\Delta\cosh{\widetilde\Delta\over 2}
\right].
\tea
After some computation we obtain the spectrum to be
\be\label{12.13.1-5}
C(k,\Sigma) = (1+h\Gamma_1)\;\coth {\pi k\over H_0},
\te
a form which shows its approximately thermal nature, with
\be\label{12.13.1-6}
\Gamma_1 = -{\widetilde\Sigma{\pi k\over H_0}\over \sinh{\pi k\over 
H_0}} + {\widetilde\Sigma\over \cosh {\pi k\over H_0}} + (\widetilde\Sigma+1)
\left[1-\left(\tanh{\pi k\over H_0}\right){2\over\pi}\int_0^\infty 
{u\sin{2ku\over H_0}\over\sinh^2u}\;du\right].
\te
As a function of $k/H_0$ the integral looks much like $\tan^{-1}$,
tending to $\pi/2$ as $k/H_0\rightarrow\infty$.

In the low frequency limit the departure from a thermal spectrum is, to 
$O(k^2)$:
\be\label{12.13.2-1}
h\Gamma_1 \simeq h\left[\widetilde\Sigma+1 - (\widetilde\Sigma + 2/3)
\left({\pi k\over H_0}\right)^2\right] \sim h\widetilde\Sigma.
\te
Note that we stipulated that $|h \widetilde \Sigma| \sim |\alpha t| \ll 1$, 
so that $h\Gamma_1$ remains small as time passes.  In the high frequency limit 
the departure is given by
\be\label{12.13.2-2}
h\Gamma_1 \rightarrow -2 h\widetilde\Sigma \;e^{-\pi k/H_0}\left({\pi k\over 
H_0}-1\right)
\te
which again remains small, and is especially close to zero for high 
frequencies.

\section{Discussion}

This paper continues the theme of our previous papers on the stochastic
approach to particle creation \cite{HM2,RHA,RHK,HuRav96} with focus on two 
main points:
1) {\it A unified approach} to treat thermal particle creation from both
spacetimes with and without event horizons based on the interpretation proposed
by one of us \cite{HuEdmonton,cgea}
that the thermal radiance can be viewed as quantum noises
of the field amplified by an exponential scale transformation
in these systems (in specific vacuum states) \cite{Dalian}.
In contradistinction to viewing these as global, geometric effects,
this viewpoint emphasizes the kinematic effect of scaling on
the vacuum.
2){\it An approximation scheme} to show that near-thermal radiation is emitted
from systems in spacetimes undergoing near-exponential expansion.
We wish to demonstrate the relative ease in constructing
perturbation theory using the stochastic approach.

The emphasis of the statistical field theory is on how quantum and 
thermal fluctuations
of the matter fields are affected by different kinematic or dynamic conditions.
For particle creation in spacetimes with event horizons,
such as for an accelerated observer and black holes, this method derives
the Hawking and Unruh effect \cite{Ang,HM2}
from the viewpoint of amplification of quantum noise and
exploits the fluctuation- dissipation relation which measures
the balance between fluctuations in the detector and dissipation in the field
\cite{RHA}. 
For spacetimes without event horizons, such as that in near-uniformly 
accelerated detectors or collapsing masses \cite{RHK}, and wide classes of
cosmological models, some studied here, one can describe them
with a single parameter measuring the deviation 
from uniformity (acceleration) or stationarity (expansion) which 
enters in the near-thermal spectrum of particle creation in all these systems.
The fact that we can understand  all thermal radiation generating processes
in these two apparently distinct classes of (cosmological and black hole)
spacetimes \cite{Dalian,HuRav96} and be able to calculate near-thermal radiance
in this and earlier papers testifies to the conceptual unity and
methodological capability of this approach.\\

\noindent {\bf Acknowledgement}
This work is supported in part by the  U S National Science Foundation
under grants PHY94-21849.
BLH acknowledges support from the General Research Board of the
Graduate School of the University of Maryland and the Dyson Visiting Professor
Fund at the  Institute for Advanced Study, Princeton. BLH and AR enjoyed
the hospitality of the physics department of the Hong Kong University of
Science and Technology where part of this work was done.
AM thanks the Australia Research Council for its financial support, while
DK likewise acknowledges support from the Australian Vice Chancellors'
Committee.

\end{document}